\documentclass[sigconf]{acmart}
\pdfoutput=1

\AtBeginDocument{%
  \providecommand\BibTeX{{%
    \normalfont B\kern-0.5em{\scshape i\kern-0.25em b}\kern-0.8em\TeX}}}

\copyrightyear{2020}
\acmYear{2020}
\setcopyright{acmlicensed}\acmConference[KDD '20]{Proceedings of the 26th ACM SIGKDD Conference on Knowledge Discovery and Data Mining}{August 23--27, 2020}{Virtual Event, CA, USA}
\acmBooktitle{Proceedings of the 26th ACM SIGKDD Conference on Knowledge Discovery and Data Mining (KDD '20), August 23--27, 2020, Virtual Event, CA, USA}
\acmPrice{15.00}
\acmDOI{10.1145/3394486.3412861}
\acmISBN{978-1-4503-7998-4/20/08}

\usepackage{float}
\usepackage{enumitem}
\usepackage[compact]{titlesec}
\usepackage[para]{footmisc}

\PassOptionsToPackage{hyphens}{url}\usepackage{hyperref}
\usepackage{graphicx}

\begin{document}
\fancyhead{}

\title[Effective Transfer Learning for Identifying Similar Medical Questions]{Effective Transfer Learning for Identifying Similar Questions: \\ Matching User Questions to COVID-19 FAQs
}
\author{Clara H.~McCreery}
\affiliation{}
\email{mccreery@cs.stanford.edu}
\authornote{Work done during internship}

\author{Namit Katariya}
\affiliation{}
\email{namit@curai.com}

\author{Anitha Kannan}
\affiliation{}
\email{anitha@curai.com}

\author{Manish Chablani}
\affiliation{EightSleep}
\authornote{Work done while at Curai}

\author{Xavier Amatriain}
\affiliation{}
\email{xavier@curai.com}

\renewcommand{\shortauthors}{McCreery et al.}

\newcommand{\fix}{\marginpar{FIX}}
\newcommand{\new}{\marginpar{NEW}}

\begin{abstract}
People increasingly search online for answers to their medical questions but the rate at which medical questions are asked online significantly exceeds the capacity of qualified people to answer them. This leaves many questions unanswered or inadequately answered. Many of these questions are not unique, and reliable identification of similar questions would enable more efficient and effective question answering schema. COVID-19 has only exacerbated this problem. Almost every government agency and healthcare organization has tried to meet the informational need of users by building online FAQs, but there is no way for people to ask their question and know if it is answered on one of these pages. While many research efforts have focused on the problem of general question similarity, these approaches do not generalize well to domains that require expert knowledge to determine semantic similarity, such as the medical domain. In this paper, we show how a double fine-tuning approach of pretraining a neural network on \emph{medical question-answer pairs} followed by fine-tuning on \emph{medical question-question pairs} is a particularly useful intermediate task for the ultimate goal of determining medical question similarity. While other pretraining tasks yield an accuracy below 78.7\% on this task, our model achieves an accuracy of 82.6\% with the same number of training examples, an accuracy of 80.0\% with a much smaller training set, and an accuracy of 84.5\% when the full corpus of medical question-answer data is used. We also describe a currently live system that uses the trained model to match user questions to COVID-related FAQs.
\end{abstract}



\maketitle

\section{Introduction}
Even before the advent of the COVID-19 pandemic, people across the world were turning to the internet to find answers to their medical
concerns \cite{pewresearch}. Around 7\%  of Google’s daily searches were health related, equivalent to around 70,000 queries every minute \cite{telegraphgoogle}. With the emergence of medical question-answering websites such as ADAM \footnote{\url{www.adam.com}}, WebMD \footnote{\url{www.webmd.com}}, AskDocs\footnote{\url{https://www.reddit.com/r/AskDocs/}} and HealthTap \footnote{\url{www.healthtap.com}}, people now 
have the opportunity to ask detailed questions and find answers, \emph{from experts}, that satisfied their needs. COVID-19 has done nothing but accelerate this trend. Almost every government agency and healthcare organization has tried to meet the informational need of users by building online FAQs that try to address as many COVID-related topics as possible (see CDC's \footnote{\url{https://www.cdc.gov/coronavirus/2019-ncov/faq.html}}, WHO's \footnote{\url{https://www.who.int/emergencies/diseases/novel-coronavirus-2019/question-and-answers-hub}}, or Mayo Clinic's \footnote{\url{https://www.mayoclinic.org/patient-visitor-guide/covid-19-faqs}} FAQs for example)


The examples above already illustrate two important problems of any medical Q\&A collection: (1) there is a very large number of possible questions that can be formulated in different ways, and (2) it is not easy for a user to browse through a large collection of pre-existing questions to find the one that most resembles their need. A scalable solution to overcome both of these issues is to build a system that can automatically match \emph{user formulated} questions with semantically similar \emph{answered} questions, and provide those as suggestions to the users. If no similar answered questions exist, we can mark them as priority for experts to respond. This approach more directly satisfies user needs allowing them to use their own words to formulate the question. It also provides an avenue for collecting unanswered questions that users want answered, which is extremely important in a rapidly changing situation such as the currrent COVID-19 pandemic.



The problem of matching general \emph{unanswered} questions with semantically similar \emph{answered} questions has been well-studied in the context of online user forums \cite{Xiaobing08, bogdanova-etal-2015-detecting, Chali18, das-etal-2016-together}, community QA \cite{Cai11learningthe, Ji12, zhang14} and question answer archives \cite{Jeon05, Ji12}.  Typical approaches either assume a large amount of training data on which, either statistics can be computed or models can be learned. However, these approaches fall short when applied to the problem of medical question similarity. First, medical questions imbibe a large amount of medical information that a single word can completely change the meaning of the question. As an example, \textit{I’m pregnant and I believe I’ve been infected with coronavirus. What should I know about going to the hospital? } and \textit{Should I visit the doctor if I am expecting and think I might have COVID-19?} are similar questions with low overlap, but \textit{Is it safe to take Vitamin D3 supplements to build immunity against Coronavirus?} and \textit{Is it safe to take Hydroxychloroquine to build immunity against Coronavirus?} are critically different and only a couple of words apart. Second, there is no publicly available medical question-question similarity data at the scale where these differences can be effectively encoded in order to learn a reliable similarity function. In fact, we hypothesize that constructing such large datasets that cover the large functional space of nuanced variations in medical domain can be quite hard, and is not a scalable proposition.


\begin{table}[ht]
\caption{Examples from our MQP dataset}
\label{tab:examples_1}
\begin{center}
\begin{tabular}{p{0.005\textwidth}p{0.15\textwidth}p{0.15\textwidth}p{0.06\textwidth}}
\hline
\multicolumn{1}{l}{}&\multicolumn{1}{l}{\bf Question 1} &\multicolumn{1}{l}{\bf Question 2} &\multicolumn{1}{c}{\bf Label} \\ 
\hline
1 & After how many hours from drinking an antibiotic can I drink alcohol?
& I have a party tonight and I took my last dose of Azithromycin this morning. Can I have a few drinks?
& Similar \\
\hline
2 & What specific exercises would help bursitis of the suprapatellar?
& Can I take any medication for pain due to suprapatellar bursitis? Unable to exercise. :(
& Different \\
\hline
3 & What does a medical physicist do during cancer treatment? & Do different types of cancers have different treatment modalities? & Different \\
\hline
\end{tabular}
\end{center}
\end{table}

\begin{figure}[ht]
\begin{center}
\includegraphics[width=0.45\textwidth]{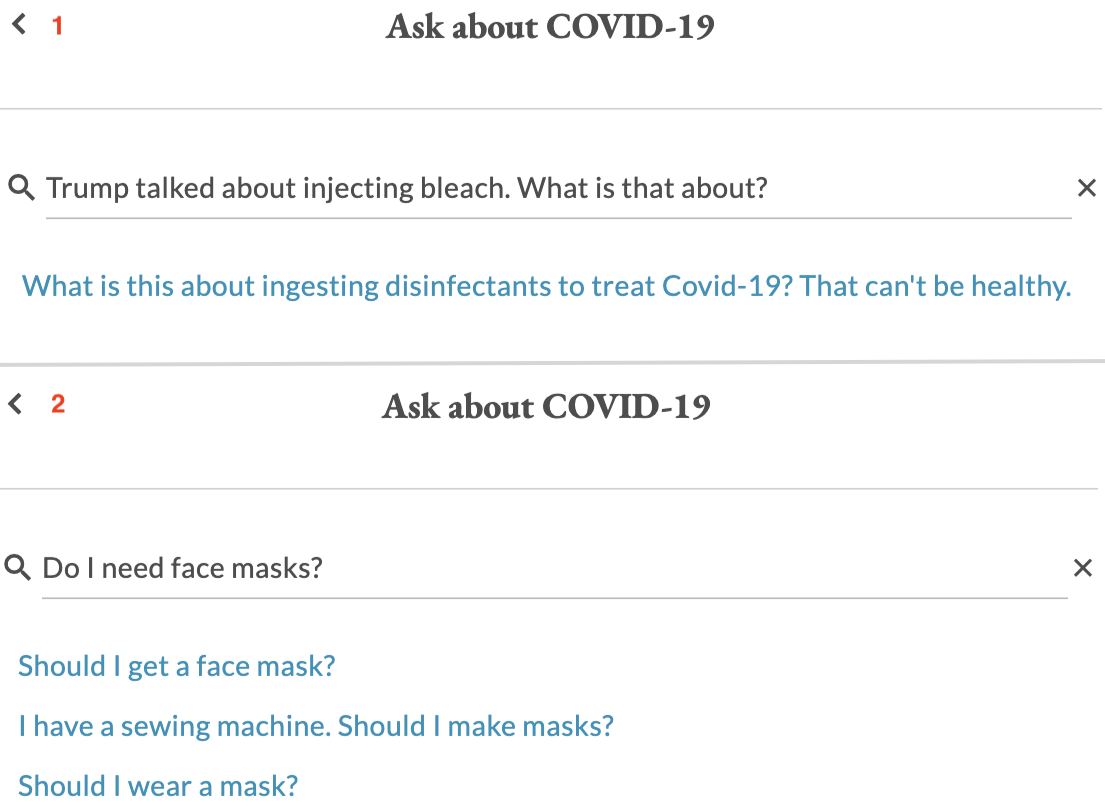}
\end{center}
\caption{Example FAQs returned by our deployed service}
\label{fig:faq_matching}
\end{figure}

In this paper, we tackle the general problem of medical question-question similarity, assuming only a small amount of labeled data of similarity pairs. We also apply the general solution to a specific COVID-19 scenario (see figure~\ref{fig:faq_matching}) where many different questions from different sources are integrated into a user-friendly experience. Our proposed solution stems from two key insights: First, whether or not two questions are semantically similar is akin to asking whether or not the answer to one also answers the other. This means that the answers in the answered questions contain wealth of medical knowledge that can be distilled into the model. The second insight is that we can infuse this medical knowledge from the answers as a pretraining task within a language model, so that we can capture relatedness between words/concepts in the language. Recent success of pretrained bi-directional transformer networks for natural language processing in non-medical fields supports this insight \cite{Peters_2018, devlin2018bert, radford2019language, yang2019xlnet, Liu_2019}.

Our approach stems from augmenting a general language model such as BERT, with medical knowledge by process of double fine-tuning that first distills medical knowledge using a large corpus of relevant in-domain task of \emph{medical question-answer pairs}. Subsequently, it fine-tunes on the available small corpus of \emph{question-question similarity dataset}. Our models pretrained on medical question-answer pairs outperform models pretrained on out-of-domain question similarity with high statistical significance. In particular, while other pretraining tasks yield an accuracy below 78.7\% on this task, our model achieves an accuracy of 82.6\% with the same number of training examples, an accuracy of 80.0\% with a much smaller training set, and an accuracy of 84.5\% when the full corpus of medical question-answer data is used.



The main contributions of this paper are:
\begin{itemize}
  \item We present an approach of double fine-tuning for the problem of question-question similarity: This helps the model to cope up with data sparsity, by imbibing domain knowledge through an intermediate fine-tuning task.  
  \item We prove that, particularly for medical NLP, domain matters: pretraining on a different task in the same domain outperforms pretraining on the same task in a different domain. However, using extensive experimentation we show that the choice of the in-domain task matters: choosing the task that provides ample signal to capture the domain knowledge needed for the final task is central.
  \item We apply the general approach medical question similarity to COVID-19 specific questions
  \item We release\footnote{\url{https://github.com/curai/medical-question-pair-dataset}} a dataset of medical question pairs generated and labeled by doctors that is based upon real, patient-asked questions, hereafter referred as \emph{MQP} dataset. Some sample examples from this dataset is provided in Table~\ref{tab:examples_1}.
\end{itemize}

The rest of the paper is structured as follows:  \S~\ref{data:qq} describes the methodology used in creating a dataset that will be made publicly available. \S~\ref{sec:approach} provides the overview of the approach. \S~\ref{sec:covid} describes how we used the model to build a service that matches user's COVID-19-related questions to FAQs published online. \S~\ref{results-section} describes experimental details and the key results,
\S~\ref{sec:related} discusses related work and we end with a discussion on future work.
\section{Medical Question Similarity Dataset for Fine-tuning}
\label{data:qq}

There is no existing dataset that we know of for medical question similarity. Therefore, one contribution of this paper is that we have generated such a dataset that we refer to as \emph{MQP} and are releasing it. This dataset is hand-generated by doctors and contains \emph{3048 medical questions pairs} that are labeled similar or different. We explicitly choose doctors for this task because determining whether or not two medical questions are the same requires medical training that crowd-sourced workers rarely have.

We present doctors with a list of 1524 patient-asked questions randomly sampled from publicly available crawl of Healthtap \citep{durakkerem}. In all of the intermediate tasks that we consider, we make sure to exclude these sampled questions. \emph{Each question results in one similar and one different pair} through the following instructions provided to the labelers:
\begin{enumerate}[topsep=2pt]
    \item Rewrite the original question in a different way while maintaining the same intent. Restructure the syntax as much as possible and change medical details that would not impact your response (ex.`I'm a 22-y-o female' could become `My 26 year old daughter' ).
    \item Come up with a related but dissimilar question for which the answer to the original question would be WRONG OR IRRELEVANT. Use similar key words. 
\end{enumerate}

The first instruction generates a \textit{positive} question pair (similar) and the second generates a \textit{negative} question pair (different). With the above instructions, we intentionally frame the task such that positive question pairs can look very different by superficial metrics, and negative question pairs can conversely look very similar. This ensures that the task is not trivial.

In Table~\ref{tab:examples_2}, we provide examples of how the doctors perform this task. Table~\ref{tab:examples_1} has more examples of the pairs they generate. 

\begin{table}[ht]
\caption{For each question in column 1, doctors are instructed to come up with a similar question (column 2) and a dissimilar question 
related to the original question (column 3). 
Columns 1 and 2 are used to generate similar question pairs while columns 1 and 3 generate dissimilar question pairs to arrive at final
dataset as in Table~\ref{tab:examples_1}}
\label{tab:datarelease} 
\begin{center}
\begin{tabular}{p{0.15\textwidth}p{0.15\textwidth}p{0.15\textwidth}} \\
\hline
\multicolumn{1}{c}{\bf Original Question} &\multicolumn{1}{c}{\bf Similar Question} &\multicolumn{1}{c}{\bf Different Question} \\ 
\hline 
If I had hepatitis a, does that mean I can't drink alcohol for a certain number of weeks afterwards?
& How soon can I drink alcohol after being tested positive for Hep A?
& Can Hep A spread via sharing cigarettes or alcohol bottles of an infected person? \\
\hline 
Am I over weight (192.9) for my age (39)?
& I am a 39 y/o male currently weighing about 193 lbs. Do you think I am overweight?
& What diet is good for losing weight? Keto or vegan? \\
\hline
What specific exercises would help bursitis of the suprapatellar?
& Hey doc! My doctor diagnosed me with suprapatellar bursitis. Are there any exercises that I can do at home?
& Can I take any medication for pain due to suprapatellar bursitis? Unable to exercise. :( \\
\hline
\end{tabular}
\end{center}
\end{table}

We  anticipate  that  each  doctor  interprets  these  instructions slightly differently, so to reduce bias, no doctor providing data in the train set generates any data in the dev or test set. Thus instead of a random train-dev-test split, the splits are created based on the doctors that labeled the examples.  In other words, we make sure that the set of doctors that created the examples in the training set is disjoint from those that created examples in the dev or test set. Furthermore, we also ensure that there is no overlap between the seed questions in the train and test set.


The final dataset contains 4567 unique questions. The minimum, maximum, median and average number of tokens in these questions are 4, 81, 20 and 22.675 respectively. 
showing there is variance in the length of the questions. The shortest question is ``Are fibroadenomas malignant?''
To obtain an oracle score, we also have doctors hand-label question pairs that a different doctor generated. The accuracy of the second doctor with respect to the labels intended by the first (viz. inter-annotator agreement) is used as an oracle and is 87.6\% in our test set of 836 question pairs. 
\section{Approach Overview}
\label{sec:approach}

We are interested in learning a model that determines whether two medical questions are similar, \emph{i.e} have semantic correspondence.
If a large corpus of pairs of similar medical questions is available, it would be relatively straightforward to learn a model 
for question similarity, as done in the case of the large-scale Quora dataset (QQP) \cite{csernai_2017, Chali18} for general question-question similarity on Quora platform.

However, labeled training data is still one of the largest barriers to supervised learning, particularly in the medical field where it is expensive to get doctor time for hand-labeling data. To overcome this issue, we take the approach of double fine-tuning, derived from transfer learning. Double fine-tuning works as follows: Starting with a pretrained model trained on a large general corpus, the model is subsequently fine-tuned \emph{twice}. In the first fine-tuning stage, a related task with large amounts of training data is used to train the model. The goal of this step is to have the model imbibe the requisite knowledge into an otherwise generic model. Our main dataset for this purpose is the medical question-answering (QA) dataset described under \textit{BERT+QA} model in section~\ref{models}, where the goal is to predict if the given answer correctly answers the given question. The final fine-tuning is performed using a small amount of labeled data available for the final goal. In our case, this refers to the task of identifying question similarity and the dataset we use is MQP, described in section~\ref{data:qq}. Both tasks are posed as binary classification problems with cross entropy loss.

\begin{figure}[ht]
\begin{center}
\includegraphics[width=0.5\textwidth]{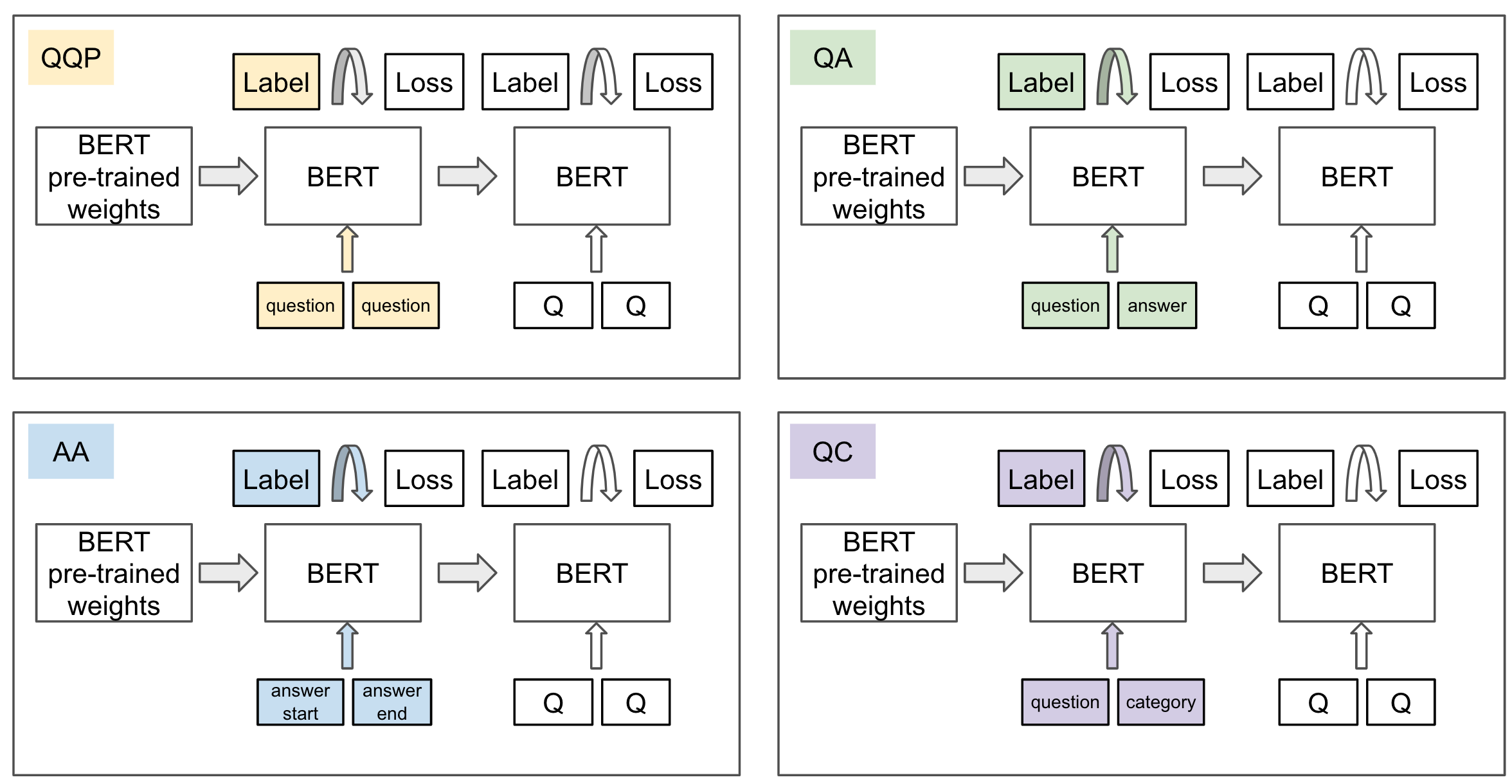}
\end{center}
\caption{We perform a double fine-tune from BERT to an intermediate task to our medical question-similarity task for four different intermediate tasks: quora question-question pairs (top left), medical question-answer pairs (top right), medical answer-answer pairs (bottom left), and medical question-category pairs (bottom right)}
\label{fig:process}
\end{figure}

In order to understand the importance of intermediate fine-tuning, we also experiment with different types of intermediate tasks.  In Figure~\ref{fig:process}, we provide the overall structure of the model training phases where each sub-figure corresponds to different intermediate tasks that we evaluate. Each of these models are described in section~\ref{models}, with details of how the intermediate tasks are setup.

For the base model for representation, we use the  architecture and weights from BERT \citep{devlin2018bert}.   we also compare against previous state-of-the-art (SOTA) models of BioBERT \citep{lee2019biobert}, SciBERT \citep{beltagy2019scibert}, and ClinicalBERT \citep{huang2019clinicalbert}. Note that these three BERT models that have been fine-tuned once already on the original BERT tasks but with different text corpora. We also perform an ablation over pretrained model architecture and reproduce our results starting with the XLNet model instead of BERT.





\section{Matching COVID-related user questions to FAQs}
\label{sec:covid}

As the coronavirus crisis has proliferated, one useful source of information has been FAQs, published by various sources such as CDC, FDA, NYTimes and others. Moreover, these FAQs have continued to evolve and are still evolving as we learn new information about the disease, prevention and safety measures. Curai Health is a remote healthcare platform where one can get treatment for many common health issues from real doctors, without having to go to a doctor's office. We deployed a service on our platform that enables users to enter their question in free-text and attempts to match their question to an existing FAQ. The goal of the system is to match a user question to a given set of question-answer pairs (FAQs).

\subsection{Problem formulation}
While the answer to a question can also be useful in knowing whether the user's question is relevant to a given FAQ, we simplified this problem to identifying questions in our FAQs that were similar to the user question, ignoring the answer. Therefore, given the pair of (user question, FAQ question), we can now use our double-finetuned BERT based question-similarity model to predict whether the questions are similar or not. 

\subsection{Inference-time data preprocessing}
Since the model was not trained on coronavirus-related data and it did not have terms such as `coronavirus' and `COVID' in its tokenizer vocabulary, the model was yielding unexpected results when the questions were input as is. Therefore, we replaced such terms in both the user question as well as the FAQs with generic placeholders like `disease'. Since the launched service made it clear that the questions were meant to be COVID-19 related and the set of FAQs pertain to the same topic as well, performing such replacements is acceptable and will not change the semantic meaning of the question.  Therefore, questions such as ``How can I protect myself from COVID-19?'' were transformed into ``How can I protect myself from the disease?'' 

\subsection{Inference}
Since we have only a few hundred curated FAQs, we score every (user question, FAQ) pair using the BERT model to get similar pairs. As we scale up to serving more FAQs, we will need to come up with an effective candidate generation scheme before running BERT inference. With our current throughput, we anticipate the current approach of scoring all FAQs to continue to work till we reach a couple of thousand FAQs.

Since our model was not fine-tuned on any kind of COVID questions or text, we found that it did make errors. Since our platform allows the users to chat with a medical professional for free, we wanted to bias towards precision than recall i.e. model predictions should actually be similar and it is acceptable to output no results even if there are actually relevant FAQs in our curated set. We achieved this by ensuring some minimum amount of ``key'' token overlap between the user question and the FAQ using a tf-idf based filter.

\subsection{Deployment}
Given the nature of the BERT model and the latency required to provide a good user experience, we used two NVIDIA Tesla K80 GPUs for inference. We encapsulate the model as a microservice, containerize it and run the container image on Google Kubernetes Engine in Google Cloud Platform.

\begin{figure}[ht]
\begin{center}
\includegraphics[width=0.45\textwidth]{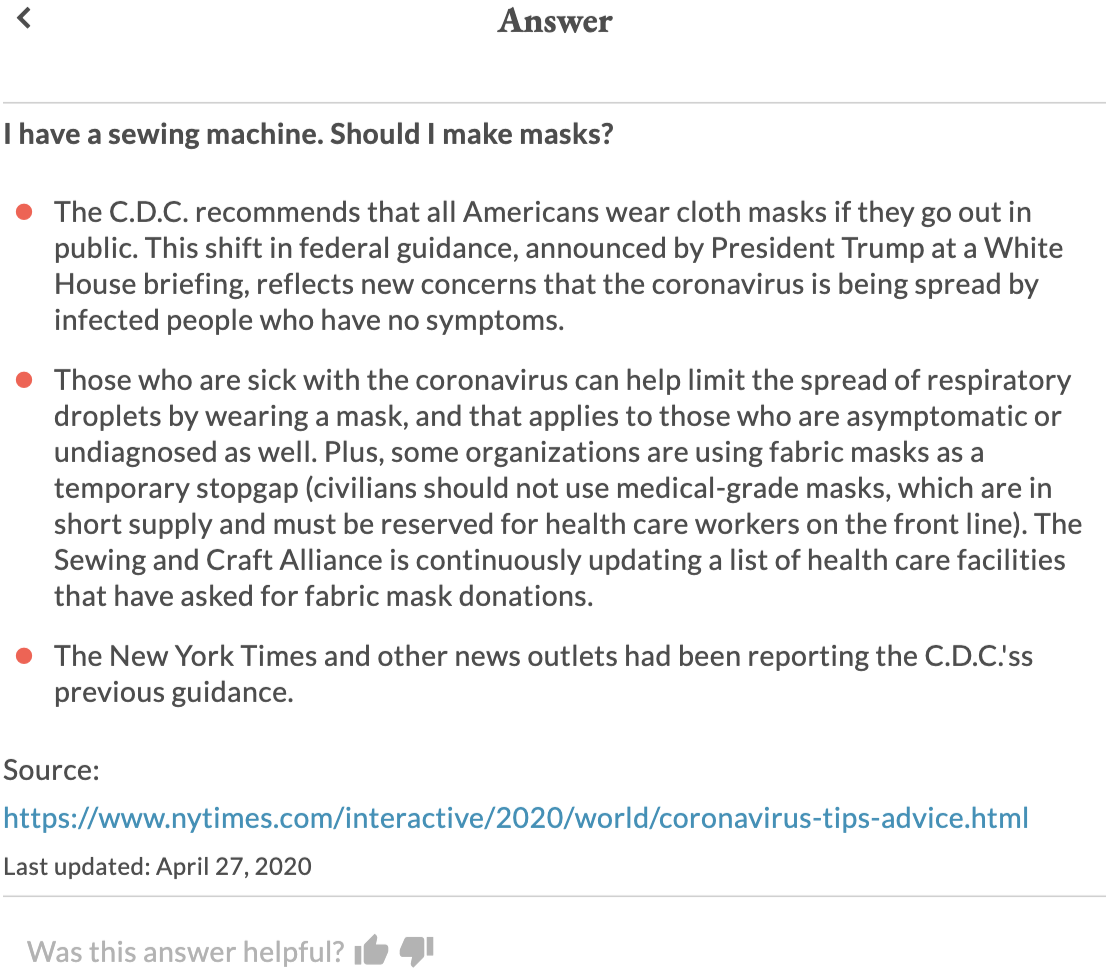}
\end{center}
\caption{Example of a rendered FAQ}
\label{fig:faq}
\end{figure}

The service can be seen in action in figure~\ref{fig:faq_matching}. As shown in figure~\ref{fig:faq}, for each FAQ that we render, we also display who it came from (source) and when it was last updated.

\section{Results}
\label{results-section}
Since (1) we did not have a readily available dataset of COVID specific questions to quantify performance on COVID question similarity and (2) the model itself is applicable for general medical question similarity, we evaluate the model performance on general medical question similarity. This section describes the datasets, models and the evaluation setup along with the results. 
\subsection{Datasets}
We used the following datasets to derive the pretraining tasks.

\textbf{Quora Question Pairs (QQP)} is a labeled corpus of 363,871 question pairs from Quora, an online question-answer forum \citep{csernai_2017}. These question pairs cover a broad range of topics, most of which are not related to medicine. However, it is a well-known dataset containing labeled pairs of similar and dissimilar questions.

\textbf{HealthTap} is a medical question-answering website in which patients can have their questions answered by doctors. We use a publicly available crawl \citep{durakkerem} with 1.6 million medical questions. Each question has corresponding long and short answers, doctor meta-data, category labels, and lists of related topics. We reduce this dataset to match the size of QQP via random sampling for direct performance comparisons, but also run one experiment leveraging the full corpus.

\textbf{WebMD} is an online publisher of medical information including articles, videos, and frequently asked questions (FAQ). For a second medical question-answer dataset, we use a publicly available crawl \citep{lasseregin} over the FAQ of WebMD with 46,872 question-answer pairs. We decrease the size of QQP and HealthTap to match this number before making direct performance comparisons.

\subsection{Models for comparison}
\label{models}
We are interested in understanding the role of double fine-tuning, and in particular the effectiveness of 
intermediate task before the final fine tuning with training set from \emph{MQP} (\S~\ref{data:qq}). For this intermediate training, we consider the following training variations (see Figure~\ref{fig:process}).

\noindent\textbf{BERT}: This is the baseline model without any intermediate training.   

\noindent \textbf{BERT+QQP}: This is BERT trained using the Quora Question Pairs dataset \cite{csernai_2017} on the binary classification task of classifying a given pair of questions as similar or dissimilar.

\noindent\textbf{BERT+QC}: 
Here, we take questions from HealthTap, pair them up with their main-category labels and call these positive examples. We then pair each question with a random other category and call this a negative example. There are 227 main categories represented, such as abdominal pain, acid reflux, acne, ADHD, alcohol etc. We then train a BERT model to classify category matches and mismatches, rather than predict to which of the classes each example belongs to.

\noindent \textbf{BERT+AA}: One task that has been known to generalize well is that of next-sentence prediction, which is one of two tasks used to train the original BERT model. To mimic this task, we take each answer from HealthTap and split it into two parts: the first two sentences (start), and the remaining sentences (end). We then take each answer start and end that came from the same original question and label these pairs as positives. We also pair each answer start with a different end from the same main category and label these as negatives. This is, therefore, a binary classification task in which the model tries to predict whether an answer start is completed by the given answer end.

\noindent\textbf{BERT+QA}: This is the proposed approach for imbibing medical knowledge into the classifier.  In order to correctly determine whether or not two questions are semantically similar, as is our ultimate goal, a network must be able to interpret the nuances of each question. Another task that requires such nuanced understanding is that of pairing questions with their correct answers. We isolate each true question-answer pair from the medical question-answering websites and label these as positive examples. We then take each question and pair it with a random answer \emph{from the same main category or tag} and label these as negative examples. Finally, we train BERT to label question-answer pairs as either positive or negative.

\begin{figure}[ht]
\begin{center}
\includegraphics[width=0.5\textwidth]{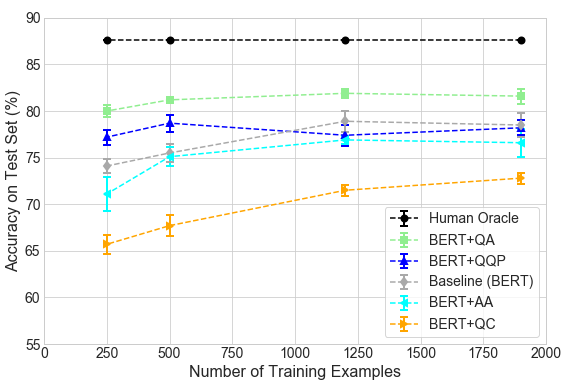}
\end{center}
\caption{The intermediate task of training on question-answer pairs (BERT+QA) reliably outperforms other intermediate tasks: Quora question pairs (BERT+QQP), medical answer completion (BERT+AA), and medical question categorization (BERT+QC). Differences are exacerbated with fewer training examples. Error bars represent one standard deviation across different data splits.}
\label{fig:results}
\end{figure}

\subsection{Experimental setup}

\noindent\textbf{Metrics:} We report accuracy as our metric. For a dataset consisting of $T$ question pairs, accuracy is defined as:
$\textrm{accuracy} = \frac{1}{T}\sum_{t=1}^{T} I[\hat{y}^{(t)} = y^{(t)}]$,
where, for $t^{th}$ example, $\hat{y}^{(t)}$ is the model's prediction of whether the pair is similar and $ y^{(t)}$ is the ground truth label, where $I$ denotes the indicator function.

\noindent\textbf{Training details:} For each intermediate task, we train the network for 5 epochs \citep{Liu_2019} with 364,000 training examples to ensure that differences in performance are not due to different dataset sizes. We then fine-tune each of these intermediate-task-models on a small number of labeled, medical-question pairs until convergence. A maximum sentence length of 200 tokens, learning rate of 2e-5, and batch size of 16 is used for all models. All experiments are done with 5 different random train/validation splits to generate error bars representing one standard deviation in accuracy. We use accuracy of each model, as described above, as our quantitative metric for comparison and a paired t-test to measure statistical significance.

\subsection{Domain Matters}
Here we investigate whether domain of the training corpus matters more than task-similarity when choosing an intermediate training step for the medical question similarity task. Accuracy on the final task (medical question similarity) is our quantitative proxy for performance.

\textbf{Domain Similarity vs Task Similarity} We fine-tune BERT on the intermediate tasks of Quora question pairs (QQP) and HealthTap question answer pairs (QA) before fine-tuning on the final task to compare performance. We find that the QA model performs better than the QQP model by $2.4\%$ to $4.5\%$, depending on size of the final training set (Figure~\ref{fig:results}). Conducting a paired t-test over the 5 data splits used for each experiment, the p-value is always less than 0.0006, so this difference is very statistically significant. We thus see with high confidence that models trained on a related in-domain task (medical question-answering) outperform models trained on the same question-similarity task but an out-of-domain corpus (QQP). Furthermore, when the full corpus of question-answer pairs from HealthTap is used, the performance climbs all the way to $84.5\% \pm 0.7\%$.

\textbf{Results hold across models} The same trends hold when the BERT base model is replaced with XLNet, with a p-value of 0.0001 (Table~\ref{tab:results}). To benchmark ourselves against existing medical models, we compare our fine-tuned models to BioBERT, SciBERT, and ClinicalBERT as the base model. Each of these models has fine-tuned the original BERT weights on a medically relevant corpus using the original BERT tasks. Given only the base model has changed, we compare BERT fine-tuned on our MQP dataset with each of these off-the-shelf models fine-tuned on MQP. All of them perform comparably to the original BERT model. Accuracies were as follows: BERT$=78.5\pm1.32$, ClinicalBERT$=74.2\pm1.72$, BioBERT$=78.5\pm0.75$ and SciBERT$=75.8\pm1.19$. We hypothesize that this is because technical literature and doctor notes that these models are pretrained on have their own distinct vocabularies. While they are more medical in nature than Wikipedia articles, they still use language quite distinct from the colloquial medical question-answer language found online.

\textbf{Results hold across datasets} We repeat our experiments with a question-answer dataset from WebMD and restrict the HealthTap and QQP dataset sizes for fair comparison. We find that the QA model again outperforms the QQP model by a statistically significant margin (p-value 0.049) and that the WebMD model even outperforms the HealthTap model with the same amount of data (Table~\ref{tab:results}). Our findings therefore hold across multiple in-domain datasets.

\begin{table}[ht]
\renewcommand\arraystretch{1.05}
\caption{Three sets of results comparing use of an in-domain question-answer task (QA) to an out-of-domain question-similarity task (QQP) for pretraining}
\label{tab:results}
\begin{center}
\begin{tabular}{lccc}
\multicolumn{1}{c}{Model} &\multicolumn{1}{c}{\bf XLNet}
&\multicolumn{1}{c}{\bf BERT}
&\multicolumn{1}{c}{\bf BERT} 
\vspace{1mm}
\\
\multicolumn{1}{c}{\begin{minipage}{0.7in}Intermediate Train Set Size\end{minipage}} &\multicolumn{1}{c}{\bf 364k}
&\multicolumn{1}{c}{\bf 364k}
&\multicolumn{1}{c}{\bf 27k}
\vspace{1mm}
\\
\hline 
Baseline \\ (No intermediate) & $77.7\% \pm 2.1\%$ & $78.5\% \pm 1.3\%$ & $ 78.5\% \pm 1.3\%$
\vspace{1mm}
\\ 
Quora Question\\ Pairs (QQP) & $78.2\% \pm 0.2\%$ & $78.2\% \pm 0.8\%$ & $77.9\% \pm 0.4\%$
\vspace{1mm}
\\ 
HealthTap (QA) & $\textbf{82.6\%} \pm 0.8\%$ & $\textbf{81.6\%} \pm 0.8\%$ & $78.3\% \pm 0.7\%$
\vspace{1mm}
\\
WebMD (QA) & -- -- & -- -- & $\textbf{79.2\%} \pm 1.2\% $
\end{tabular}
\end{center}
\end{table}

\subsection{Not All In-Domain Tasks Embed Relevant Medical Knowledge}

We investigate further the extent to which task matters for an in-domain corpus in two different ways. We start by using the same HealthTap data and forming different tasks from the questions therein, and then we compare our models against intermediate models trained by other researchers.

To test the extent to which any in-domain task would boost the performance of an out-of-domain model, we compare \textbf{BERT+QA} to \textbf{BERT+AA} and \textbf{BERT+QC}. As before, we use accuracy on the final question-similarity task as our proxy for performance and keep the test set constant across all models. Figure~\ref{fig:results} shows the results. We find that both of these tasks actually perform worse than the baseline BERT model, making final model less useful for understanding the subtler differences between two questions. We hypothesize that for \textbf{BERT+QC}, it is very easy for questions being in the same category to be dissimilar and therefore it is likely that it hasn't learned useful question representations for question similarity. Similarly for \textbf{BERT+AA}, the language in answers can be  different than the personal language used in patient-asked questions so that the learned representations for question language might not be ideal. This suggests that while domain does matter a lot, many tasks are not well-suited to encoding the proper domain information from the in-domain corpus. 

\subsection{Qualitative Analysis}
To get a better qualitative understanding of performance, we perform an error analysis on our trained models. We define a consistent error as one that is made by at least four of the five models trained on different train/validation splits. Similarly, we consider a model as getting an example consistently correct if it does so on at least four of the five models trained on different train/validation splits.  

\begin{table}[ht]
\caption{Examples that were consistently labeled wrong by at least one model type. Patterns reveal the key differences in what is learned by each intermediate task}
\label{fig:additionalerrors}
\begin{center}
\includegraphics[width=0.47\textwidth]{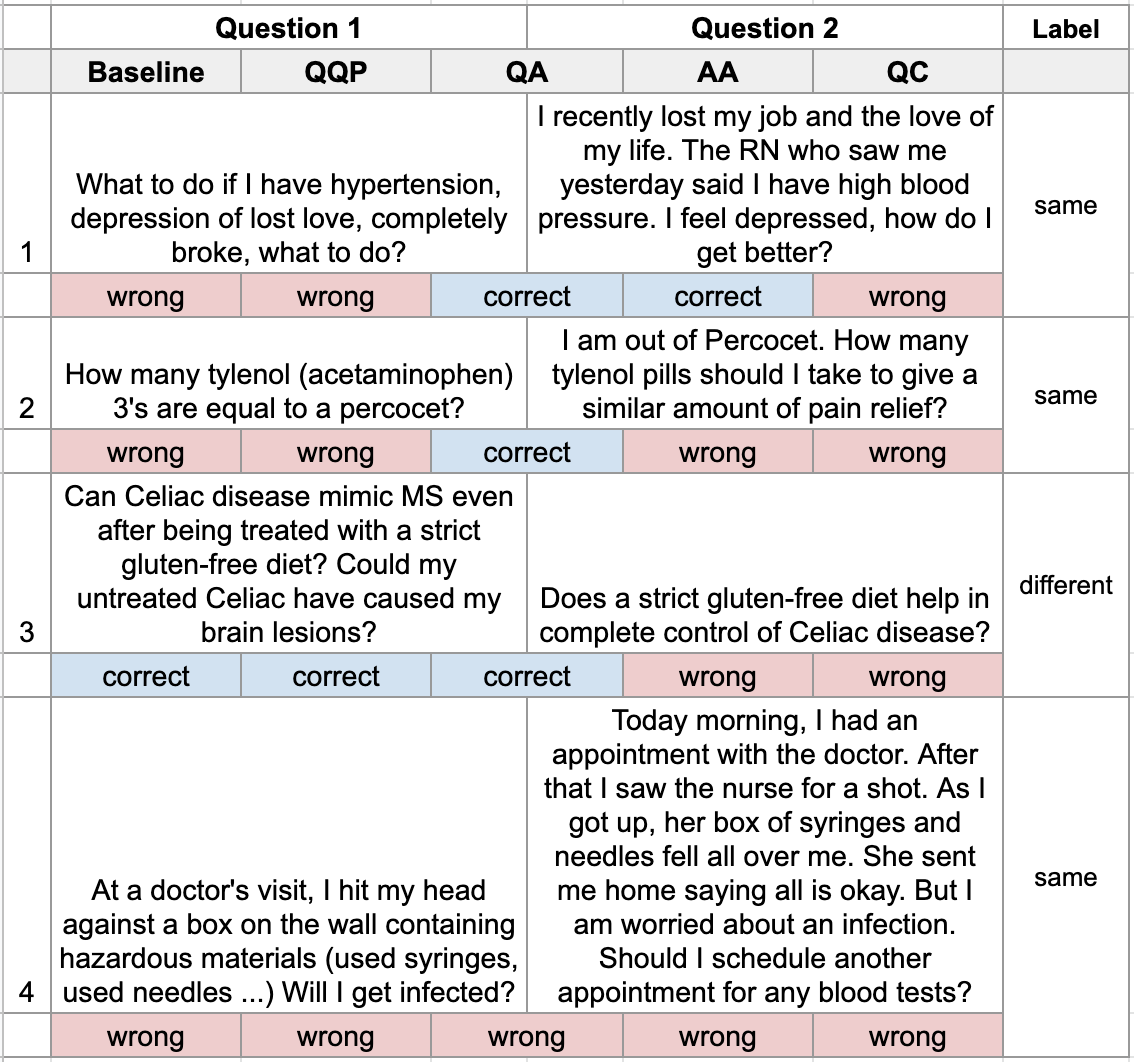}
\end{center}
\end{table}

By investigating the question pairs that a model-type gets consistently wrong, we can form hypotheses about why the model may have failed on that specific example. Table~\ref{fig:additionalerrors} shows examples of pairs of questions with their true label and how each of the models labeled the pair. We form hypotheses based on key pieces of medical language that we could potentially perturb to make the models not make the mistakes that they do. For instance, from row 1 in table~\ref{fig:additionalerrors}, we hypothesize that the nuance is hypertension being a synonym for high blood pressure. Other than QA and AA, rest of the models don't get it right because they would not have seen enough medical domain training instances that highlight this equivalence.
However, there need not always be well-encapsulated and only a handful of concepts that can be tweaked. Row 4 shows an example of fairly complex ways of formulating similar questions that one can't easily distill to a few edits to make them understandable for the models.

\textbf{An approach to understanding model errors:} We can prove or disprove our hypotheses by augmenting each question pair to add or remove one challenging aspect of the language, at a time and observe whether or not those changes result in a different label. We repeatedly make such small changes to the input until the models label those examples correctly. Note that the augmented questions are not added to our test set and do not contribute to our quantitative performance metrics; they are only created for the sake of probing and understanding the trained models .

We demonstrate our approach in Table~\ref{fig:erroranalysis} by showing our analysis for one specific question pair. Question 1 is kept fixed and we incrementally make small changes to Question 2 to inspect model predictions.
\footnote{Note that the \textit{"mixed"} label implies that at most 3 out of the 5 models trained on each of the 5 random splits labeled the example correctly while a "correct" / "wrong" implies that at least 4 out of those 5 models consistently labeled them "correct" / "wrong"} 
While not shown here, using the same approach, we find that differences in spelling and capitalization do not cause a significant number of errors in any model, although they are present in many questions.

We believe that this analysis not only helps us shed some light on explainability of our models but also guides the collection of additional training data through strategies such as active learning.

\begin{table}[ht]
\caption{Example question pair that was augmented to reveal which aspects of the question pair the network failed to understand}
\label{fig:erroranalysis}
\begin{center}
\includegraphics[width=0.49\textwidth]{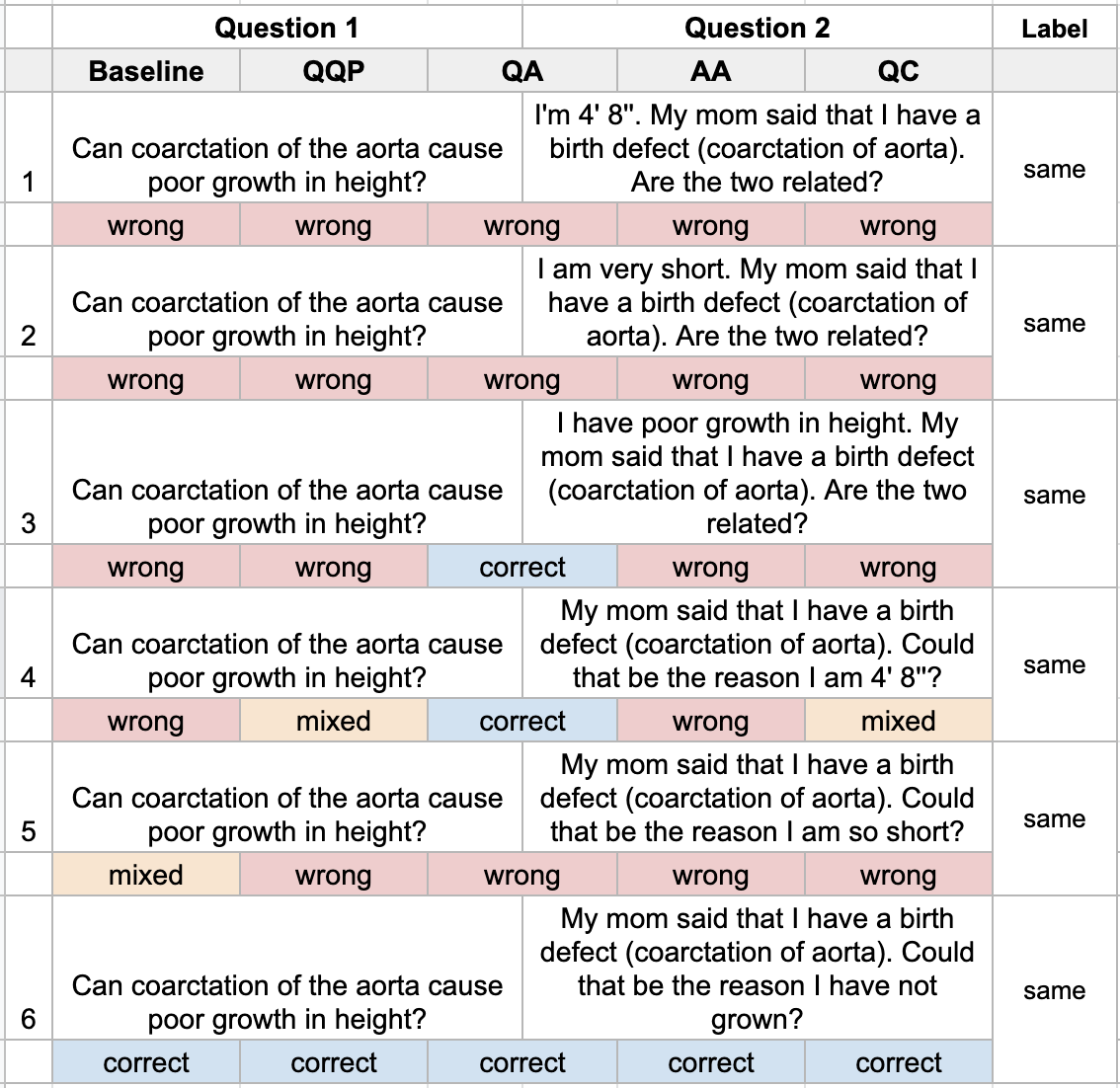}
\end{center}
\end{table}
\section{Related Work}
\label{sec:related}
\noindent {\bf Medical Question-Question similarity}

While there has been significant work on question-question similarity (\cite{bogdanova-etal-2015-detecting,  lei_2016} and references therein), 
research in medical question-question similarity is still somewhat nascent.  The closest to our work is that of \citet{abacha2016, abacha2019questionentailment}. We differ from this work in significant ways. First, rather than training a model to answer medical questions correctly, we train a model to determine if any existing questions in the dataset is semantically similar to the new question. Second, unlike their approach that constructs training data using specialized rules and manual curation, we use the approach of transfer learning where a surrogate in domain task with large amounts of data is used to infer medical knowledge, and subsequently fine-tune on a small corpus of manually labeled question similarity pairs. Finally, we are interested in questions that are patient-asked which tend to use less technical language, include more misspellings, and span a different range of topics than the language and distribution of pairs that they generate using FAQs. \\


\noindent {\bf Generation of medical question pairs}

Previous work has tried to overcome this using augmentation rules to generate similar question pairs automatically \citep{li2018finding}, but this leads to an overly simplistic dataset in which negative question-pairs contain no overlapping keywords and positive question-pairs follow similar lexical structures. Another technique for generating training data is weak supervision \citep{Ratner_2017}, but due to the nuances of determining medical similarity, generating labeling functions for this task is difficult. 

Also related to medical question pairs generation is that of the problem of recognizing question entailment (RQE) \citep{MEDIQA-RQE}. While question entailment allows for 
asymmetric similarity metric where one question is more specific than the other, question similarity requires that the metric is symmetric. Second, as pointed out previously, their question pairs have a different language and topic distribution from the one we care about. \\

\noindent {\bf Pretrained Networks for General Language Understanding}

NLP has undergone a transfer learning revolution in the past year, with several large pretrained models earning state-of-the-art scores across many linguistic tasks. Two such models that we use in our own experiments are BERT \citep{devlin2018bert} and XLNet \citep{yang2019xlnet}. These models have been trained on semi-supervised tasks such as predicting a word that has been masked out from a random position in a sentence, and predicting whether or not one sentence is likely to follow another. The corpus used to train BERT was exceptionally large (3.3 billion words), but all of the data came from BooksCorpus and Wikipedia. \citet{talmor2019multiqa} recently found that BERT generalizes better to other datasets drawn from Wikipedia than to tasks using other web snippets. This is consistent with our finding that pretraining domain makes a big difference.\\

\noindent {\bf Double Fine-tuning for Domain Transfer}

To address the need for pretrained models in particular domains, some researchers have recently re-trained BERT on different text corpora such as scientific papers \citep{beltagy2019scibert}, doctor's medical notes \citep{huang2019clinicalbert} and biomedical journal articles \citep{lee2019biobert}. However, re-training BERT on the masked-language and next-sentence prediction tasks for every new domain is unwieldy and time-consuming. We investigate whether the benefits of retraining on a new domain can also be realized by fine-tuning BERT on other in-domain tasks. \citet{phang2018sentence} see a boost with other tasks across less dramatic domain changes, where a different text corpus is used for the final task but not an entirely different technical vocabulary or domain.
\section{Conclusions and Future Work}
\label{future}

In this work, we release MQP, a dataset of medical question pairs generated and labeled by doctors that is based upon real, patient-asked questions. We also show that the double finetuning approach of pretraining on in-domain question-answer matching (QA) is particularly useful for the difficult task of identifying semantically similar questions. Furthermore, we show that the choice of this in-domain task matters: choosing a task that provides ample signal to capture the domain knowledge is needed to be able to perform the final task well.

Although the QA model outperforms the out-of-domain same-task QQP model, there are a few examples where the QQP model seems to have learned information that is missing from the QA  model. In the future, we can further explore whether these two models learned independently useful information from their pretraining tasks. If they did, then we hope to be able to combine these features into one model with multi-task learning. An additional benefit of the error analysis is that we have a better understanding of the types of mistakes that even our best model is making. It is therefore now easier to use weak supervision and augmentation rules or even active learning to supplement our datasets to increase the number of training examples in those difficult regions of the data. Both of these improvements could further improve our performance on this task.


\bibliographystyle{ACM-Reference-Format}
\bibliography{acmart}


\end{document}